\documentclass[]{iopart}

\usepackage{graphicx,color,setstack,iopams,a4wide}

\usepackage{cite}

\begin{document}

\title{Silicane and germanane: tight-binding and first-principles studies}

\author{V.\ Z\'{o}lyomi, J.\ R.\ Wallbank, and V.\ I.\ Fal'ko}

\address{Physics Department, Lancaster University, Lancaster LA1 4YB,
  United Kingdom}

\ead{v.zolyomi@lancaster.ac.uk}

\date{\today}

\begin{abstract} We present a first-principles and tight-binding model
study of silicane and germanane, the hydrogenated derivatives of
two-dimensional silicene and germanene. We find that the materials are stable in freestanding
form, analyse the orbital composition, and derive a tight-binding
model using first-principles calculations to fit the parameters.
\end{abstract}

\submitto{2D Materials}


\maketitle


Recent years have seen a rapid rise in research interest towards atomically thin two-dimensional (2D) materials.
Graphene has been in the focus of intensive research since its discovery \cite{novoselov_2004,geim_2007},
followed by silicene, Si$_2$ \cite{Silicene01,Silicene02,Silicene03,DrummondSilicene}.
Practical application of graphene in nanoelectronics is somewhat limited by the lack of a band gap,
and while alternatives such as hexagonal boron nitride \cite{BN01,BN02} and transition metal dichalcogenides
\cite{mos2exfol,KisA_2011_2,KisA_2011,WS2exp01,WS2exp02,AtacaC_2012,Morpurgo} exist and others
have been predicted theoretically such as gallium chalcogenides \cite{ZolyomiGaX}, significant research effort has been
put into engineering a band gap in graphene.
Complete hydrogenation was shown to be another way to engineer a band gap in graphene. The resulting
material, graphane \cite{Graphane} (C$_2$H$_2$) has a buckled honeycomb structure with a single hydrogen atom attached to each
carbon site on alternating sides of the sheet. A recent experiment \cite{Germanane} has shown that few-layer
germanane (Ge$_2$H$_2$), the hydrogenated germanene, can be synthesised,
expanding the family of atomic 2D materials.

In this work we provide a first-principles study of silicane and germanane.
We present the phonon dispersions to illustrate that these materials are dynamically
stable.
The band structures, effective masses, charge carriers, and an orbital
decomposition of the valence and conduction bands at high symmetry points in the Brillouin zone are presented. The latter
information is utilised for building a tight-binding model for the description of the valence and conduction
bands of silicane and germanane.

\begin{figure}
\begin{center}
\includegraphics[clip,scale=0.24,angle=0]{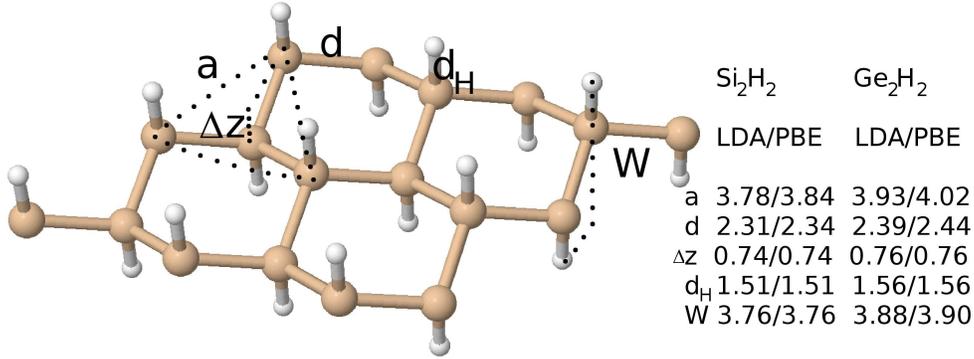}
\caption{Structure of silicane and germanane. The larger spheres denote Si/Ge atoms while the smaller
ones denote hydrogens. Here $a$ is the lattice constant, $d$ the
bond length between neighbouring Si/Ge atoms, $\Delta z$ is the sublattice buckling between the A and B sublattice, $d_{H}$
is the distance between Si/Ge and the hydrogen, and $W$ is the total width of the layer (excluding
van der Waals radii of hydrogens). The structural paremeters are given in units of \AA.
\label{fig:structure}
}
\end{center}
\end{figure}

\label{SecHamiltonian}

(I) Silicane and germanane have a honeycomb lattice as depicted in Fig. \ref{fig:structure}. The unit cell comprises two Si/Ge
atoms and two H atoms, and the A and B sublattices of Si/Ge atoms exhibit a buckling.
The tight-binding model we employ is an all-valence second-nearest neighbour model that takes into account four (one $s$ and
three $p$) electrons per
Si/Ge atom and the single electron of hydrogen. The tight-binding Hamiltonian is

\begin{equation}
\fl
\begin{array}{lll}
H&=&H_{0}+H_{1}+H_{2}\\
H_0&=&\sum_{i}(\varepsilon_{s} a^{+}_{i} a_{i} +\varepsilon_{p} \sum_{\alpha} (b^{+}_{i\alpha} b_{i\alpha})+
\varepsilon_{s^{H}} c^{+}_{i} c_{i})\\
H_1&=&\sum_{i}(\gamma_{s^{H}s} c^{+}_{i} a_{i})+\gamma_{ss} a^{+}_{A} a_{B}+
\sum_{\alpha} (\gamma_{sp} v_{\alpha}^{AB} a^{+}_{A} b_{B\alpha}+
\gamma_{pp\sigma}(v_{\alpha}^{AB})^{2}  b^{+}_{A\alpha} b_{B\alpha}+\gamma_{pp\pi}(1-\\
&&(v_{\alpha}^{AB})^{2}) b^{+}_{A\alpha} b_{B\alpha})+
\sum_{\alpha\neq\beta} (\gamma_{pp\sigma}v_{\alpha}^{AB}v_{\beta}^{AB} b^{+}_{A\alpha} b_{B\beta}-
\gamma_{pp\pi}v_{\alpha}^{AB}v_{\beta}^{AB} b^{+}_{A\alpha} b_{B\beta})+\\
&&\sum_{\alpha} \gamma_{s^{H}p} v_{\alpha}^{AB} c^{+}_{A} b_{B\alpha}
+h.c.\\
H_2&=&\sum_{i}(
\gamma^{'}_{ss} a^{+}_{i} a_{i'}+
\sum_{\alpha} (\gamma^{'}_{sp} v_{\alpha}^{AA'} a^{+}_{i} b_{i'\alpha}+
\gamma^{'}_{pp\sigma}(v_{\alpha}^{AA'})^{2}  b^{+}_{A\alpha} b_{A'\alpha}+\gamma^{'}_{pp\pi}(1-\\
&&(v_{\alpha}^{AA'})^{2}) b^{+}_{A\alpha} b_{A'\alpha})
+\sum_{\alpha\neq\beta} (\gamma^{'}_{pp\sigma}v_{\alpha}^{AA'}v_{\beta}^{AA'} b^{+}_{A\alpha} b_{A'\beta}-
\gamma^{'}_{pp\pi}v_{\alpha}^{AA'}v_{\beta}^{AA'} b^{+}_{A\alpha} b_{A'\beta})+\\
&&\gamma^{'}_{s^{H}s} c^{+}_{A} a_{B}+
\sum_{\alpha} \gamma^{'}_{s^{H}p} v_{\alpha}^{AB'} c^{+}_{A} b_{B\alpha}
+h.c.\\
v_{\alpha}^{AB}&=&({\mathbf{R}_{Si/Ge}^{A}}-{\mathbf{R}_{Si/Ge}^{B}})_{\alpha}/|{\mathbf{R}_{Si/Ge}^{A}}-{\mathbf{R}_{Si/Ge}^{B}}|\\
v_{\alpha}^{AA'}&=&({\mathbf{R}_{Si/Ge}^{A}}-{\mathbf{R}_{Si/Ge}^{A'}})_{\alpha}/|{\mathbf{R}_{Si/Ge}^{A}}-{\mathbf{R}_{Si/Ge}^{A'}}|,
v_{\alpha}^{AB'}=({\mathbf{R}_{Si/Ge}^{A}}-{\mathbf{R}_{H}^{B'}})_{\alpha}/|{\mathbf{R}_{Si/Ge}^{A}}-{\mathbf{R}_{H}^{B'}}|\end{array}
\nonumber
\end{equation}

{\noindent Here, $a^{+}$ and $a$ are the creation and annihilation operators of the $s$ electrons of Si/Ge, 
$b^{+}$ and $b$ are the same for the $p$ electrons of Si/Ge, while $c^{+}$ and $c$ are the same for the electrons of
the H atoms.
In $H_0$, parameters $\varepsilon_{s}$ and $\varepsilon_{p}$ are the on-site energies of the $s$ and $p$ orbitals of Si/Ge,
$\varepsilon_{s^{H}}$ is the on-site energy of the electron of the hydrogen atom.
In $H_1$, $\gamma_{s^{H}s}$ parameterises the nearest neighbour hopping between $s$ orbitals
of Si/Ge and hydrogen, while $\gamma_{ss}$, $\gamma_{sp}$, $\gamma_{pp\sigma}$, and $\gamma_{pp\pi}$ are
the nearest neighbour hoppings between Si/Ge electrons on sublattice A and sublattice B.
In $H_2$, parameters $\gamma^{'}_{ss}$, $\gamma^{'}_{sp}$, $\gamma^{'}_{pp\sigma}$, and $\gamma^{'}_{pp\pi}$
are the second-nearest neighbour hoppings between
the Si/Ge electrons on the same sublattice, while $\gamma^{'}_{s^{H}s}$ and $\gamma^{'}_{s^{H}p}$ are the second-nearest neighbour
hoppings between the orbitals of Si/Ge and hydrogen on different sublattices.
Summations in $i$ go over the A and B sublattices while summations in $\alpha$ and $\beta$ go over $x,y,z$;
$v_{\alpha}^{AB}$,$v_{\alpha}^{AA'}$, and $v_{\alpha}^{AB'}$ take into account the orientation of the $p$ orbitals, where
$\mathbf{R}$ denotes the coordinates of the atoms. The Hamiltonian
in the Slater--Koster approach \cite{SlaterKoster1954},
shown with solid lines separating A and B sublattice contributions as well as
hydrogen contributions has the form of a $10 \times 10$ matrix:}

\begin{equation}
\fl
\begin{array}{l}
\left(\begin{array}{cccc|cccc|cc}
H_{ss}^{'} & H_{sx}^{'} & H_{sy}^{'} & H_{sz}^{'} & H_{ss} & H_{sx} & H_{sy} & H_{sz} & H_{s^{H}s} & H_{s^{H}s}^{'}\\
H_{sx}^{'*} & H_{xx}^{'} & H_{xy}^{'} & H_{xz}^{'} & H_{sx} & H_{xx} & H_{xy} & H_{xz} &  & H_{s^{H}x}^{'}\\
H_{sy}^{'*} & H_{xy}^{'*} & H_{yy}^{'} & H_{yz}^{'} & H_{sy} & H_{xy} & H_{yy} & H_{yz} &  & H_{s^{H}y}^{'}\\
H_{sz}^{'*} & H_{xz}^{'*} & H_{yz}^{'*} & H_{zz}^{'} & H_{sz} & H_{xz} & H_{yz} & H_{zz} & H_{s^{H}p} & H_{s^{H}z}^{'}\\
\hline H_{ss}^{*} & H_{sx}^{*} & H_{sy}^{*} & H_{sz}^{*} & H_{ss}^{'} & H_{sx}^{'} & H_{sy}^{'} & H_{sz}^{'} & H_{s^{H}s}^{'} & H_{s^{H}s}\\
H_{sx}^{*} & H_{xx}^{*} & H_{xy}^{*} & H_{xz}^{*} & H_{sx}^{'*} & H_{xx}^{'} & H_{xy}^{'} & H_{xz}^{'} & H_{s^{H}x}^{'}\\
H_{sy}^{*} & H_{xy}^{*} & H_{yy}^{*} & H_{yz}^{*} & H_{sy}^{'*} & H_{xy}^{'*} & H_{yy}^{'} & H_{yz}^{'} & H_{s^{H}y}^{'}\\
H_{sz}^{*} & H_{xz}^{*} & H_{yz}^{*} & H_{zz}^{*} & H_{sz}^{'*} & H_{xz}^{'*} & H_{yz}^{'*} & H_{zz}^{'} & H_{s^{H}z}^{'} & H_{s^{H}p}\\
\hline H_{s^{H}s}^{*} &  &  & H_{s^{H}p}^{*} & H_{s^{H}s}^{'*} & H_{s^{H}x}^{'*} & H_{s^{H}y}^{'*} & H_{s^{H}z}^{'*} & H_{s^{H}s^{H}}^{'}\\
H_{s^{H}s}^{'*} & H_{s^{H}x}^{'*} & H_{s^{H}y}^{'*} & H_{s^{H}z}^{'*} & H_{s^{H}s}^{*} &  &  & H_{s^{H}p}^{*} &  & H_{s^{H}s^{H}}^{'}\end{array}\right)\\
H_{ss}=\gamma_{ss}\sum_{B}e^{i\mathbf{k}(\mathbf{\mathbf{R}_{Si/Ge}^{A}}-\mathbf{\mathbf{R}_{Si/Ge}^{B}})},
H_{s^{H}s}=\gamma_{s^{H}s},
H_{s^{H}p}=\gamma_{s^{H}p}\\
H_{s\alpha}=\gamma_{sp}\sum_{B}v_{\alpha}^{AB}e^{i\mathbf{k}(\mathbf{\mathbf{R}_{Si/Ge}^{A}}-\mathbf{\mathbf{R}_{Si/Ge}^{B}})},
H_{s\alpha}^{'}=\gamma_{sp}^{'}\sum_{A}v_{\alpha}^{AA'}e^{i\mathbf{k}(\mathbf{\mathbf{R}_{Si/Ge}^{A}}-\mathbf{\mathbf{R}_{Si/Ge}^{A'}})}\\
H_{\alpha\alpha}=\gamma_{pp\sigma}\sum_{B}(v_{\alpha}^{AB})^{2}e^{i\mathbf{k}(\mathbf{\mathbf{R}_{Si/Ge}^{A}}-\mathbf{\mathbf{R}_{Si/Ge}^{B}})}+\gamma_{pp\pi}\sum_{B}(1-(v_{\alpha}^{AB})^{2})e^{i\mathbf{k}(\mathbf{\mathbf{R}_{Si/Ge}^{A}}-\mathbf{\mathbf{R}_{Si/Ge}^{B}})}\\
H_{\alpha\beta}=\gamma_{pp\sigma}\sum_{B}v_{\alpha}^{AB}v_{\beta}^{AB}e^{i\mathbf{k}(\mathbf{\mathbf{R}_{Si/Ge}^{A}}-\mathbf{\mathbf{R}_{Si/Ge}^{B}})}-\gamma_{pp\pi}\sum_{B}v_{\alpha}^{AB}v_{\beta}^{AB}e^{i\mathbf{k}(\mathbf{\mathbf{R}_{Si/Ge}^{A}}-\mathbf{\mathbf{R}_{Si/Ge}^{B}})},\alpha\neq\beta\\
H_{\alpha\alpha}^{'}=\varepsilon_{p}+\gamma_{pp\sigma}^{'}\sum_{A}(v_{\alpha}^{AA'})^{2}e^{i\mathbf{k}(\mathbf{\mathbf{R}_{Si/Ge}^{A}}-\mathbf{\mathbf{R}_{Si/Ge}^{A'}})}+\gamma_{pp\pi}^{'}\sum_{A}(1-(v_{\alpha}^{AA'})^{2})e^{i\mathbf{k}(\mathbf{\mathbf{R}_{Si/Ge}^{A}}-\mathbf{\mathbf{R}_{Si/Ge}^{A'}})}\\
H_{\alpha\beta}^{'}=\gamma_{pp\sigma}^{'}\sum_{A}v_{\alpha}^{AA'}v_{\beta}^{AA'}e^{i\mathbf{k}(\mathbf{\mathbf{R}_{Si/Ge}^{A}}-\mathbf{\mathbf{R}_{Si/Ge}^{A'}})}-\gamma_{pp\pi}^{'}\sum_{A}v_{\alpha}^{AA'}v_{\beta}^{AA'}e^{i\mathbf{k}(\mathbf{\mathbf{R}_{Si/Ge}^{A}}-\mathbf{\mathbf{R}_{Si/Ge}^{A'}})},\alpha\neq\beta\\
H_{s^{H}s}^{'}=\gamma_{s^{H}s}^{'}\sum_{B}e^{i\mathbf{k}(\mathbf{\mathbf{R}_{Si/Ge}^{A}}-\mathbf{\mathbf{R}_{H}^{B'}})},
H_{s^{H}s^{H}}^{'}=\varepsilon_{s}^{H}+\gamma_{s^{H}s^{H}}^{'}\sum_{A}e^{i\mathbf{k}(\mathbf{\mathbf{R}_{Si/Ge}^{A}}-\mathbf{\mathbf{R}_{Si/Ge}^{A'}})}\\
H_{s^{H}\alpha}^{'}=\gamma_{s^{H}p}^{'}\sum_{B}v_{\alpha}^{AB'}e^{i\mathbf{k}(\mathbf{\mathbf{R}_{Si/Ge}^{A}}-\mathbf{\mathbf{R}_{H}^{B'}})},
H_{ss}^{'}=\varepsilon_{s}+\gamma_{ss}^{'}\sum_{A}e^{i\mathbf{k}(\mathbf{\mathbf{R}_{Si/Ge}^{A}}-\mathbf{\mathbf{R}_{Si/Ge}^{A'}})}
\end{array}
\nonumber
\end{equation}

The total number of parameters in this model is sixteen, but one can choose one of the on-site energies to be zero to set the Fermi level
leaving fifteen parameters to fit. The resulting model can be used to provide a simple semiempirical reproduction of first-principles band
structures, as we will show below.

\label{SecDFT}

(II) The parameters in this model, as well as the justification for the choice of orbitals
comes from first-principles studies using density functional theory (DFT) implemented in
the \textsc{vasp} \cite{vasp} plane-wave-basis code. First, we
calculated the lattice parameters of silicane and germanane
with multiple semilocal exchange-correlation functionals: the local density approximation (LDA)
and the Perdew-Burke-Ernzerhof \cite{pbe} (PBE) functionals. In addition the screened
Heyd-Scuseria-Ernzerhof 06 (HSE06) functional \cite{hse} was used to obtain the electronic
band structures to compensate (at least partially)
for the underestimation of the band gap by semilocal functionals.
The plane-wave cutoff energy was 500 eV\@. A $12\times 12$ Monkhorst-Pack
\textbf{k}-point grid was used for geometry optimisations while a $24\times 24$ grid was
used to calculate the band structures.
The vertical separation of periodic images of the monolayer was set to 15~\AA\@. The
force tolerance in the optimisation was 0.005 eV/\AA\@.
Phonons were calculated with the force constant approach in a $3\times 3$ supercell.

\label{Results}

We find that the relaxed structure of both silicane and germanane is very similar to that of 
graphane, as illustrated in Fig. \ref{fig:structure}. The parameters listed were obtained
after a full geometry optimisation. 
The bond lengths obtained with the PBE functional are systematically larger than those
optimised with the LDA, as expected \cite{FavotF_1999}. Note that the hydrogenation
is accompanied by a significant increase in the magnitude of the sublattice buckling when compared
to silicene and germanene, where, according to LDA/PBE, $\Delta z=0.44/0.45$~\AA~and 
$\Delta z=0.65/0.69$~\AA, respectively.
Lattice constants and sublattice bucklings agree with previous literature within $\pm 5$\%
and $\pm 10$\%, respectively \cite{SihGehLit01,SihGehLit02,SihGehLit03,SihGehLit04,SihGehLit05}.

\begin{figure}
\begin{center}
\includegraphics[clip,scale=0.55]{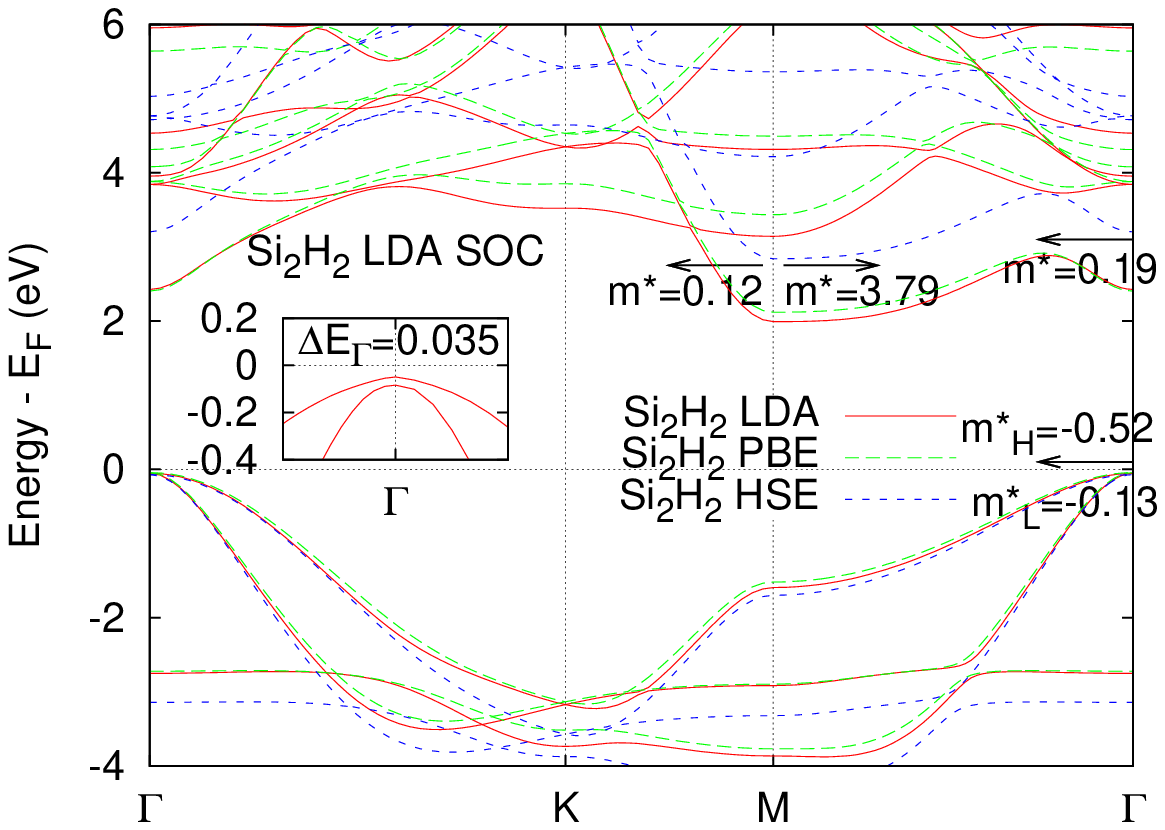}
\includegraphics[clip,scale=0.55]{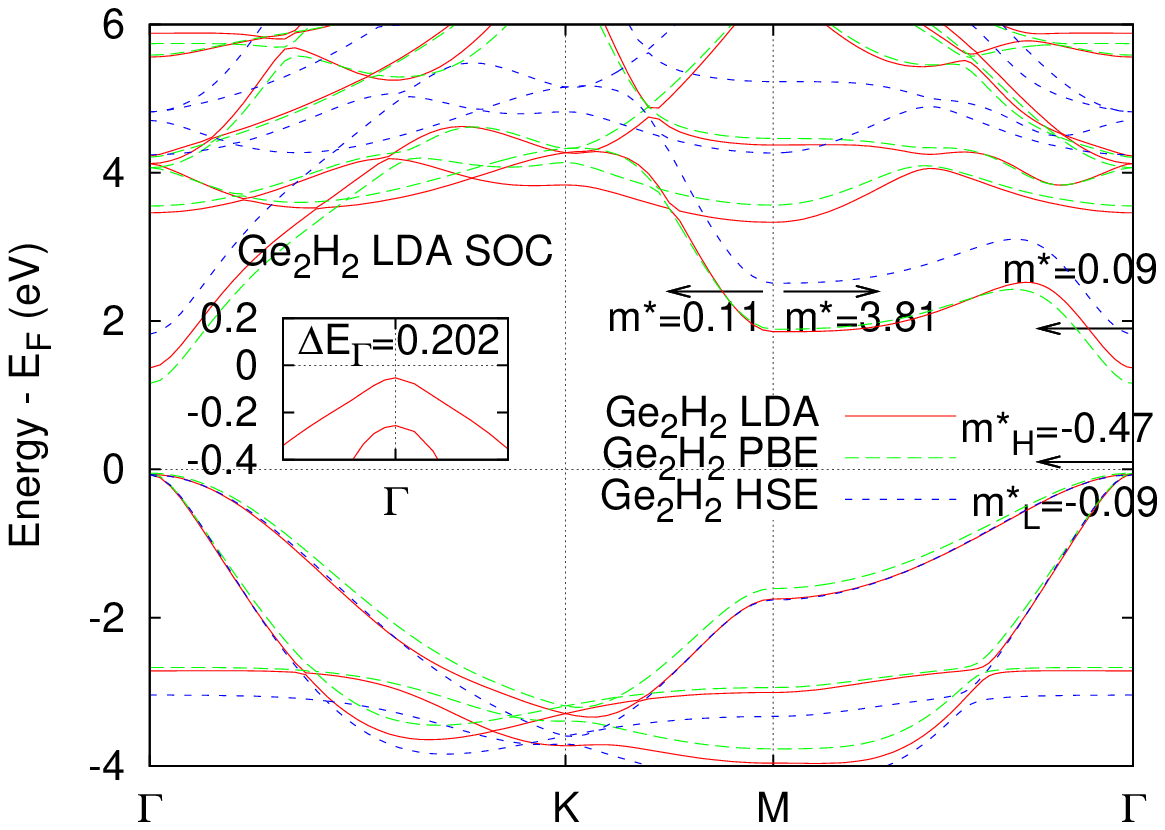}
\caption{Band structures of silicane and germanane. The zero of energy
is taken to be the Fermi level and the top of the valence band is marked with a horizontal line.
The effect of spin-orbit coupling at the $\Gamma$ point is illustrated in the insets.
Effective masses (in units of electron mass) in the HSE06 calculations are provided in the conduction band at M and $\Gamma$, and
in the valence band at $\Gamma$ (where the $H$ and $L$ subscript refers to the heavy and light effective mass).
Our results on $Ge_{2}H_{2}$ compare well with Ref. \cite{Germanane}.
It is worth noting that we find almost no sign of anisotropy in the effective masses
at $\Gamma$. In comparison to the literature on graphene, an LDA study found a small anisotropy
in both the valence and conduction band of graphane \cite{LiuaL_2009}, while an earlier GGA study makes
no mention of any such anisotropy \cite{SofoJO_2007}.
\label{fig:bands}
}
\end{center}
\end{figure}

The calculated electronic band structures are plotted in Fig. \ref{fig:bands}.
In comparison to graphane \cite{SofoJO_2007} one important difference
in silicane and germanane is that in the latter two materials a band appears close to the
conduction band edge at the M point. While the conduction band minimum of Ge$_2$H$_2$ is at the $\Gamma$
point similar to C$_2$H$_2$, in the case of Si$_2$H$_2$ it is in fact at the M point, making silicane an
indirect gap semiconductor. The band gaps of silicane and germanane are 2.91 eV and 1.90 eV, respectively, according
to the HSE06 functional which is expected to underestimate the gap by no more than 10\ \% \cite{hse10percent}.
Note that the conduction
band is anisotropic at the M point with a heavy effective mass in the $\Gamma$ direction.
Our finding that the band gap of Si$_2$H$_2$ is indirect and that of Ge$_2$H$_2$ is direct is
supported by previous literature using a variety of methods ranging from semi-local DFT through hybrid
functionals to single-shot GW \cite{SihGehLit01,SihGehLit02,SihGehLit04,SihGehLit05}.

We have also looked at the effects of spin-orbit coupling. The insets in Fig. \ref{fig:bands} show
the spin-orbit induced splitting of the valence band at the $\Gamma$ point. The effect of spin-orbit coupling
is more pronounced in the case of germanane where the valence band splits by an energy difference of 202 meV at $\Gamma$.

Now we discuss the orbital composition of the valence and conduction bands of silicane and germanane (see
Table \ref{table:orb_decomp}). We find that at
the $\Gamma$ point the valence band consists of Si/Ge $p_x$ and $p_y$ orbitals, while the conduction band is dominantly
Si/Ge $s$ and $p_z$. However, at the M point the conduction band also
contains Si/Ge $p_x$ and $p_y$ contributions. The H $s$ orbital also contributes at the M point, to the valence band in silicane
(and slightly to the conduction band in germanane). This means that for a tight-binding description of silicane and germanane
an all valence description is required taking into account the $s$, $p_x$, $p_y$, and $p_z$ orbitals of Si/Ge, as well as
the H $s$ orbital.

\begin{table}
\caption{Orbital decomposition of the valence and conduction bands of silicane and germanane at the $\Gamma$ and M points
according to the local density approximation.
\label{table:orb_decomp}}

\begin{center}
\begin{tabular}{lcc}

\hline
\hline
&$\Gamma$&M\\

\hline

Si$_2$H$_2$ val.&$0.23(p^{Si}_x+p^{Si}_y)$&$0.05p^{Si}_x+0.16p^{Si}_y$\\

Si$_2$H$_2$ cond.&$0.09s^{Si}+0.05p^{Si}_z+0.03s^{H}$&$0.07s^{Si}+0.01p^{Si}_x+0.01p^{Si}_y+0.03p^{Si}_z$\\

Ge$_2$H$_2$ val.&$0.32(p^{Ge}_x+p^{Ge}_y)$&$0.07p^{Ge}_x+0.22p^{Ge}_y$\\

Ge$_2$H$_2$ cond.&$0.24s^{Ge}+0.06p^{Ge}_z$&$0.11s^{Ge}+0.03p^{Ge}_x+0.03p^{Ge}_z$\\

\hline
\hline

\end{tabular}
\end{center}
\end{table}

\begin{figure}
\begin{center}
\includegraphics[clip,scale=0.60]{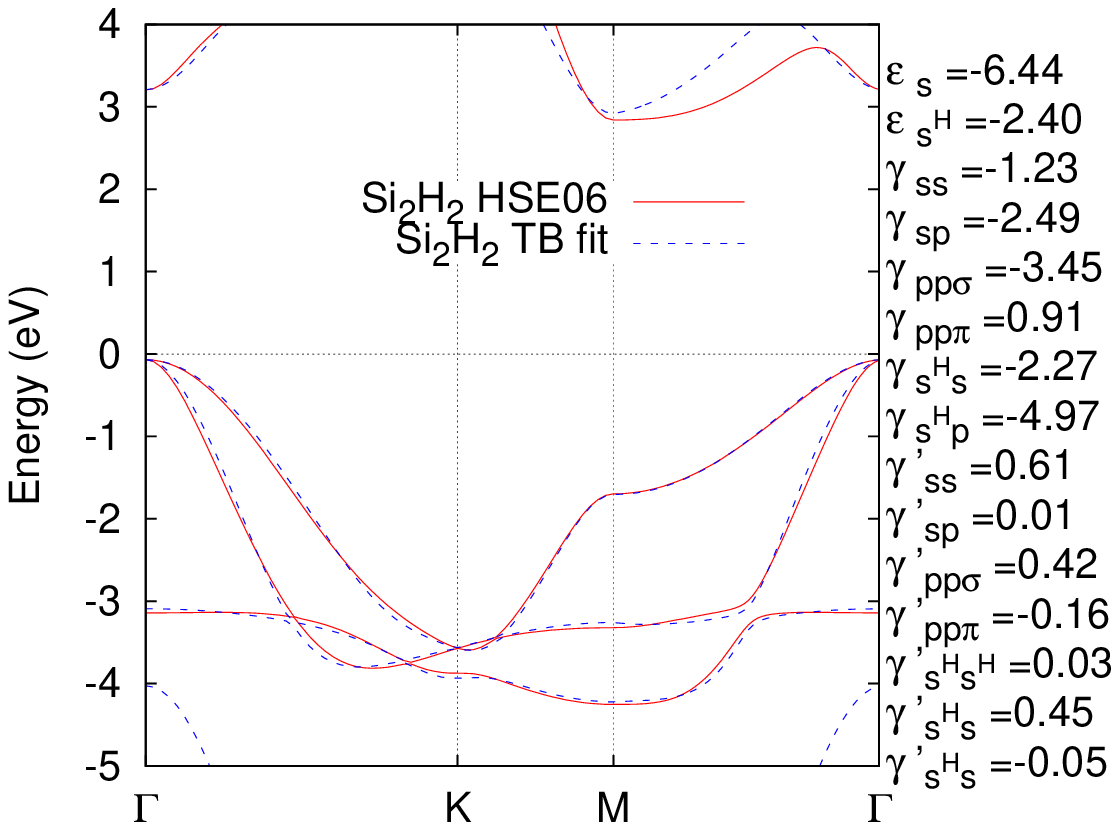}
\includegraphics[clip,scale=0.60]{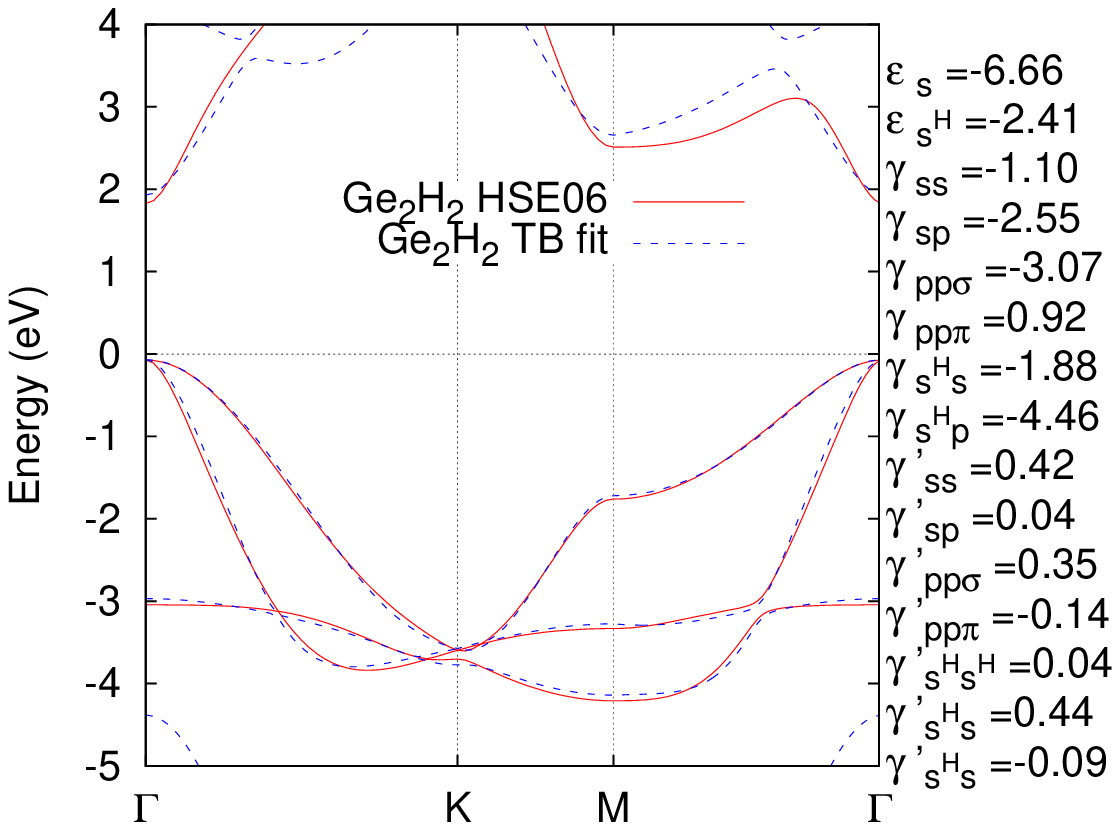}
\caption{Tight-binding band structures of silicane and germanane
compared with the HSE06 DFT bands. The parameters of the model are shown in the legend
in units of eV. The reference energy level is set by $\varepsilon_{p}=0$.
\label{fig:TBbands}
}
\end{center}
\end{figure}

(III) We used the HSE06 band structures as reference to
obtain the tight-binding parameters.  We find that the tight-binding band structure can
reproduce the entirety of the DFT valence band and the vicinity of the conduction band at both the $\Gamma$ 
and M points (see Fig. \ref{fig:TBbands}). It is important to note here that if we neglect second-nearest neighbour interactions
the valence band can still be
reproduced but the behaviour of the conduction band at the M point cannot, which indicates that the second nearest neighbour
interactions are responsible for the minimum in the conduction band at the M point.
Also, the $d$-shell of Si/Ge is likely to affect states in the conduction band. The best fit is achieved with the
parameters listed in the legend of Fig. \ref{fig:TBbands}; the fitting was optimised to give a quantitative description of the valence
band and the conduction band near the $\Gamma$ and the M point.

\label{SecStability}

\begin{figure}
\begin{center}
\includegraphics[clip,scale=0.55]{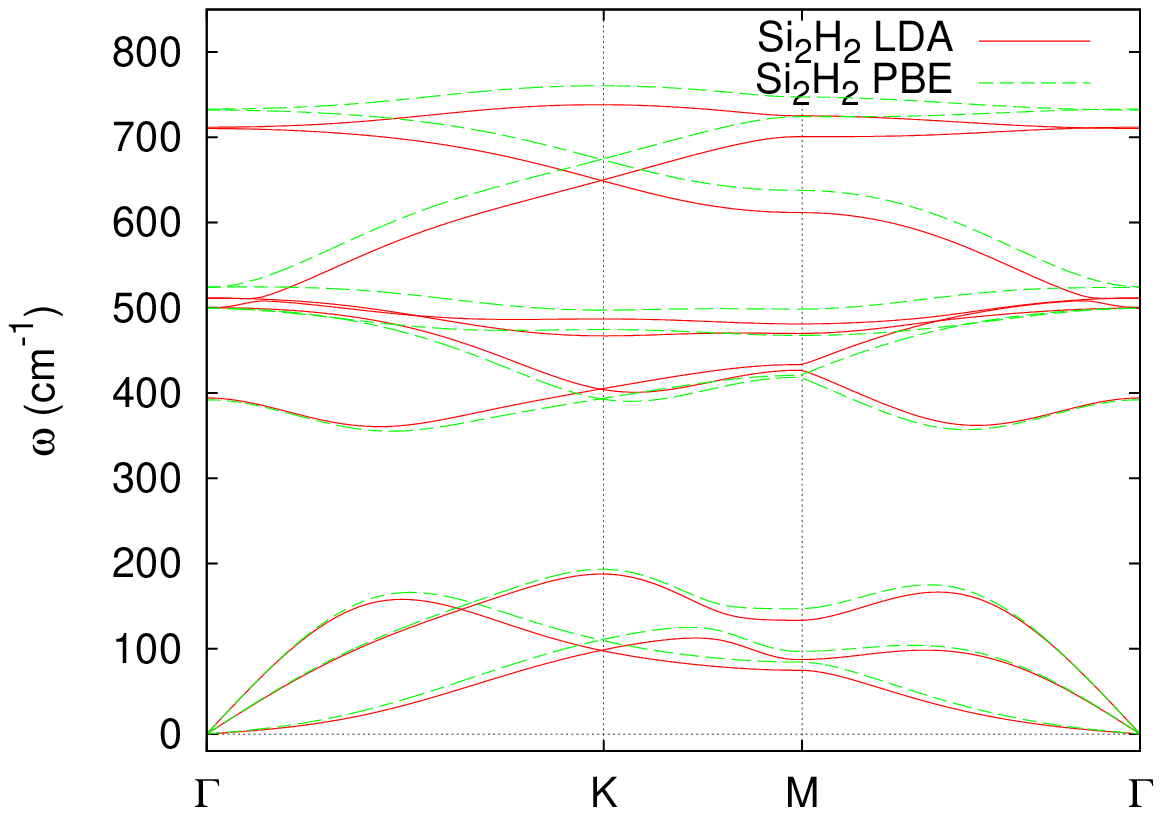}
\includegraphics[clip,scale=0.55]{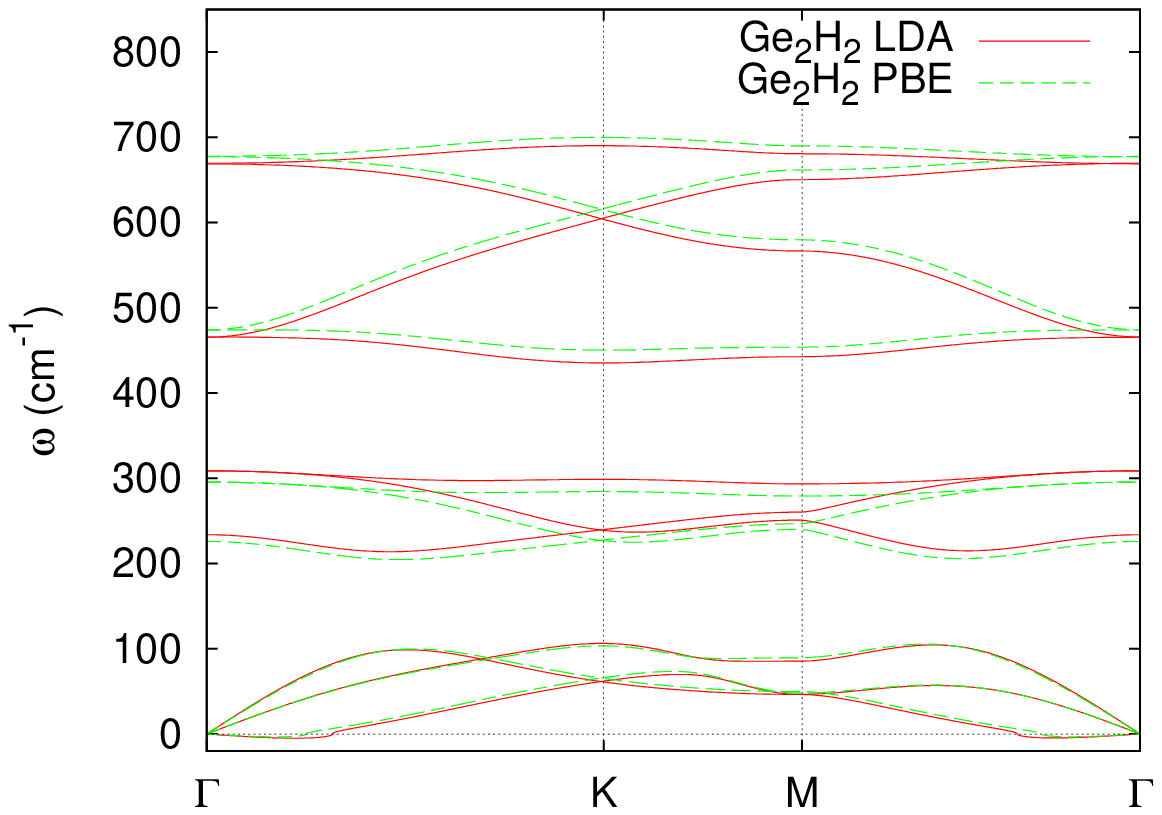}
\caption{Phonon dispersions of silicane and germanane. The high frequency branch
above 2000 cm$^{-1}$ corresponding to Si--H/Ge--H vibrations is omitted for sake of clarity.
\label{fig:phonons}
}
\end{center}
\end{figure}

(IV) Finally, we have performed a full geometry optimisation of Si$_2$H$_2$ and Ge$_2$H$_2$.
While the geometry optimisation yields an energetically stable configuration for both materials,
it is necessary to examine their phonon dispersions in order to ascertain whether they are dynamically stable.
We find that silicane is stable as there is no sign of any dynamical instability anywhere along the high
symmetry lines of the Brillouin zone (see Fig. \ref{fig:phonons}). In the case of Ge$_2$H$_2$ we find a small pocket of
instability for the flexural
acoustic phonons. We believe that this is an artifact arising due to the difficulty in converging the flexural
acoustic branch of two-dimensional materials as the phonon wave vector goes to zero. Since we find no other pockets of instability in any of the other
branches we conclude that germanane is also dynamically stable. This is an important finding as the experiments
in Ref. \cite{Germanane} were performed on multilayers of germanane on a substrate, while our calculations predict that
suspended single-layer germanane would be stable, too.

It is worth noting that there is an alternate configuration for silicane and germanane not considered here.
The so-called chair-like structure we studied corresponds to the case when H atoms alternate on the two sides of
the sheet such that for each sublattice the H atom is on a fixed side. In the so-called boat configuration
the H atoms alternate in pairs instead, which slightly increases the unit cell size. The latter has been shown to be
notably less stable than the chair configuration in the case of graphane \cite{SofoJO_2007}, nevertheless the
boat configuration of silicane and germanane has been found to be stable \cite{SihGehLit02} which is important
to bear in mind.

\label{SecConclusion}

In conclusion we have shown using first-principles density functional theory that freestanding hydrogenated silicene and germanene,
better known as silicane and germanane, are energetically and dynamically stable.
We have shown that silicane is an indirect and germanane a direct gap semiconductor and derived a tight binding model to describe
the valence and conduction bands of these materials, fitting the parameters to hybrid density functional calculations. A minimum in the
conduction band at the M point is caused by second nearest neighbour interaction which, in the case of silicane, leads to an indirect
band gap.

\ack

We acknowledge financial support from
EC-FET European Graphene Flagship Project,
EPSRC Science and Innovation Award, ERC Synergy Grant ``Hetero2D,''
the Royal Society Wolfson Merit Award, and the Marie Curie project
CARBOTRON.

\end{document}